\documentclass[journal]{IEEEtran}

\usepackage{cite}
\usepackage{amsmath,amssymb,amsfonts}
\usepackage{graphicx}
\usepackage{url}
\usepackage{textcomp}
\usepackage{xcolor}
\usepackage{booktabs}
\usepackage{multirow}
\usepackage{array}
\usepackage{pifont}
\usepackage{float}
\usepackage{tikz}
\usetikzlibrary{shapes,arrows,positioning}

\floatstyle{ruled}
\newfloat{algorithm}{t}{loa}
\newcounter{algline}
\newcommand{\algline}[1]{\stepcounter{algline}\makebox[0.28in][r]{\scriptsize\thealgline}\hspace{0.08in}#1\par}

\title{RaStream: Edge-Deployable Streaming Human Mesh Recovery from mmWave Radar}

\author{Jiazhen~Dong and Lei~Liu,~\IEEEmembership{Senior Member,~IEEE}
\thanks{Jiazhen Dong and Lei Liu are with the Zhejiang Provincial Key Laboratory of Information Processing, Communication and Networking, College of Information Science and Electronic Engineering, Zhejiang University, Hangzhou, 310007, China (e-mail: \{dongjiazhen, lei\_liu\}@zju.edu.cn).}
\thanks{Corresponding author: Lei Liu.}}

\begin{document}

\maketitle

\begin{abstract}
Millimeter-wave (mmWave) radar provides a privacy-preserving sensing modality for
human motion analysis, yet continuous SMPL eXpressive (SMPL-X) recovery on edge
devices remains difficult because accurate reconstruction benefits from dense volumetric
radar observations that preserve weak articulated-body reflections, while repeatedly
processing such high-dimensional data over time incurs substantial computational cost.
Temporal context is also necessary to resolve frame-local ambiguity, but the relevant
SMPL-X attributes evolve on different time scales, with body morphology remaining
relatively stable and articulated pose, root orientation, and global translation changing
rapidly, which makes a single temporal update mechanism suboptimal for streaming
reconstruction.
We present RaStream, a causal streaming framework that separates rich volumetric
spatial perception from lightweight temporal reasoning. For each radar observation, a
localization-conditioned spatial encoder first exploits global scene context to estimate a
coarse body location and then focuses detailed reconstruction on a center-guided region
of interest while preserving the native 3D radar structure; the resulting global and local
features are fused into a compact spatial representation, so that high-dimensional
volumetric evidence is extracted only once per observation and temporal history can
subsequently be propagated through compact representations and fixed-size persistent
states. To further account for the heterogeneous temporal dynamics of SMPL-X
attributes, RaStream maintains a slow morphology state for shape-related information
and a fast motion state for pose, orientation, and translation.
On the 661K-frame M4Human benchmark, RaStream reduces MVE from 90.90 mm for
RT-Mesh to 84.27 mm with spatial-only inference and further to 72.05 mm with causal
temporal refinement, while the Base configuration requires 26.93 ms per FP32 invocation
on a Jetson Orin Nano. These results demonstrate that continuous radar mesh recovery
can be made practical for edge streaming by extracting rich volumetric evidence once and
propagating temporal information thereafter in a compact causal form. Our code is
available at \url{https://github.com/LeiLiu-s-Lab/RaStream.git}.
\end{abstract}


\section{Introduction}
\label{sec:introduction}

\begin{figure}[!t]
\centering
\includegraphics[width=\columnwidth]{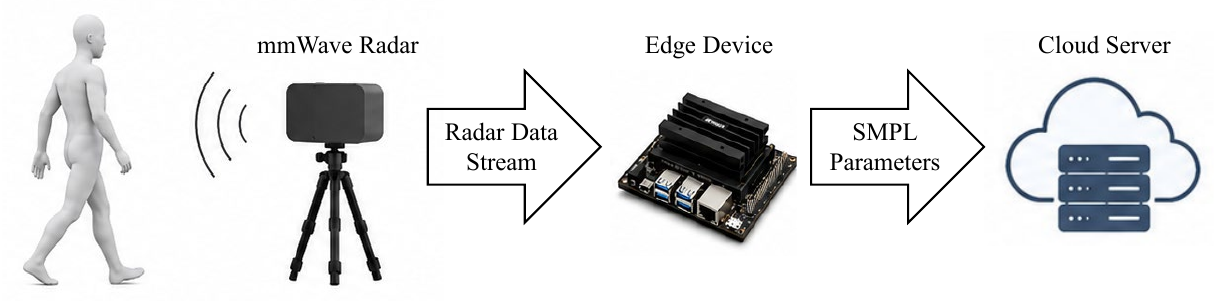}
\caption{Edge streaming scenario for privacy-preserving mmWave human mesh recovery on resource-constrained devices.}
\label{fig:scenario}
\end{figure}

Continuous human motion sensing supports a wide range of indoor monitoring and human--computer interaction applications~\cite{atzori2010internet,rashid2016applications,wang2020fall,zhang2019human}. Camera-based pose and mesh recovery has advanced rapidly~\cite{kanazawa2018hmr,kolotouros2019spin,kocabas2020vibe,li2021hybrik,zhang2021pymaf,yuan2022glamr}, but identifiable visual observations limit its use in privacy-sensitive environments. Millimeter-wave (mmWave) radar provides a complementary sensing modality that is less affected by illumination and partial visual occlusion while avoiding direct capture of visual appearance. As illustrated in Fig.~\ref{fig:scenario}, our target setting is causal mesh recovery from a continuous radar stream on a resource-constrained edge device. Prior RF/mmWave systems have addressed through-wall sensing~\cite{adib2015capturingwifi,zhao2018through}, skeletal pose estimation~\cite{sengupta2020mm,cui2023milipoint,zhao2018rf}, fine-grained radar signal generation~\cite{li2024sbrf}, gait and identity recognition~\cite{huang2023hdnet}, cross-category action recognition~\cite{sheng2025mmzear}, gesture interaction~\cite{liu2020longrangegesture}, and vital-sign monitoring~\cite{xu2021vitals}. Recent work has further extended radar sensing to real-time continuous activity recognition~\cite{liu2025zuma}, while radar benchmarks increasingly support dense human reconstruction with SMPL~\cite{loper2015smpl,zhang2022mmbody} and SMPL-X~\cite{pavlakos2019smplx,fan2026m4human}. Unsupervised learning has also been explored for general radar representations~\cite{zhang2025umimo}. These advances demonstrate the growing capability of radar-based human sensing, but they primarily target signal generation, recognition, representation learning, or frame-wise reconstruction. Continuous radar mesh recovery is a different problem: a practical streaming system must preserve weak spatial evidence, exploit temporal continuity, and satisfy causal edge-inference constraints simultaneously.

The central difficulty lies in reconciling spatial fidelity, temporal context, and streaming efficiency. Point-cloud methods such as mmMesh~\cite{xue2021mmmesh}, MiliPoint~\cite{cui2023milipoint}, mmBody~\cite{zhang2022mmbody}, and P4Transformer~\cite{fan2021p4transformer} reduce computation through peak extraction or point-level processing. Video-based synthesis can generate radar training data for privacy-preserving mobile human sensing~\cite{deng2024midaspp}, but it does not remove the need to preserve informative evidence in measured radar observations. Peak selection is computationally attractive, yet it can irreversibly discard weak responses from limbs, partially occluded body parts, and other regions that remain informative for dense articulated reconstruction. Evidence that may be redundant for coarse localization can still constrain detailed mesh recovery. Radar-tensor methods such as HuPR~\cite{lee2023hupr}, RT-Pose~\cite{ho2024rtpose}, RETR~\cite{yataka2024retr}, and RT-Mesh~\cite{fan2026m4human} preserve denser spatial evidence, but repeatedly processing full radar volumes becomes expensive in a continuous stream. Related data-generation work targets rich radar observations for generalized gesture recognition~\cite{deng2025g3r}, rather than full-body SMPL-X recovery. At the same time, frame-wise estimates remain sensitive to body orientation, self-occlusion, weak or missing returns, multipath, and clutter, making temporal context important for stable reconstruction. Existing real-time continuous radar sensing has primarily focused on activity recognition~\cite{liu2025zuma}; short-window motion models~\cite{pham2026two} provide only bounded context, whereas buffering longer radar sequences increases memory use. Recurrently processing volumetric radar features, for example with recurrent units derived from LSTM~\cite{hochreiter1997lstm}, further adds high-dimensional computation and memory traffic, while bidirectional or offline models rely on future observations unavailable during causal inference. The key question is therefore not how to process more radar data over longer windows, but how to extract rich spatial evidence once from each observation and then propagate temporal information in a compact form.
A further requirement arises from the temporal heterogeneity of SMPL-X attributes. Causal history aggregation can already stabilize frame-local radar ambiguity, but body shape and gender remain approximately stable within a subject sequence and benefit from evidence accumulated over time, whereas articulated pose, root orientation, and global translation change rapidly and must remain responsive to the current observation. A single temporal state therefore faces conflicting update requirements: rapid updates can make morphology estimates sensitive to transient radar noise, while aggressive smoothing can suppress genuine motion changes. Separating slowly varying morphology information from rapidly varying motion information is therefore expected to provide an additional refinement beyond generic causal temporal modeling.

These observations suggest three design principles for continuous mmWave mesh recovery. First, the spatial encoder should preserve the native volumetric representation while separating global body localization from subject-centered detailed reconstruction, so that expensive fine-grained processing is concentrated where it is most useful. Second, each radar observation should undergo high-dimensional spatial processing only once, after which its information should be compressed into a compact representation for temporal propagation. Third, temporal reasoning should distinguish slowly varying morphology from rapidly varying motion so that long-term evidence accumulation does not compromise responsiveness to current observations.

Based on these principles, we propose RaStream, a unified framework for spatial encoding, causal temporal refinement, and edge-aware streaming. For each radar observation, a localization-conditioned spatial encoder preserves the native 3D radar structure and first exploits global scene context to estimate a coarse body center. Detailed feature extraction is then focused on a center-guided region of interest, and gated global--local fusion compresses the resulting evidence into a compact spatial token. This token serves as the interface between spatial perception and temporal reasoning, allowing rich volumetric information to be extracted once from the current observation and temporal history to be propagated thereafter in a compact form. Causal history aggregation supplies the main temporal benefit over spatial-only inference. RaStream further maintains two persistent temporal states with different roles: a morphology state accumulates slowly varying evidence about body shape and related attributes, while a motion state tracks rapidly changing pose, orientation, and translation, providing an additional refinement beyond a single recurrent state. In this way, RaStream avoids both long raw-volume histories and recurrent processing of high-dimensional radar volumes, while retaining fixed-size causal temporal memory suitable for edge deployment.

\begin{table*}[!t]
\caption{Comparison with representative RF/mmWave sensing, reconstruction, and radar-perception methods.}
\label{tab:related_work_comparison}
\centering
\footnotesize
\setlength{\tabcolsep}{2.4pt}
\renewcommand{\arraystretch}{1.05}
\begin{tabular}{p{0.145\textwidth} p{0.215\textwidth} p{0.165\textwidth} p{0.195\textwidth} p{0.14\textwidth}}
\toprule
Method & Primary Task & Input & Output & Temporal Mode \\
\midrule
RF-Capture~\cite{adib2015capturingwifi} & Human figure sensing & RF waveform & Coarse human figure & Offline sensing \\
mm-Pose~\cite{sengupta2020mm} & Pose estimation & Radar point cloud & 3D keypoints & Frame-wise \\
HuPR~\cite{lee2023hupr} & Pose estimation benchmark & Radar tensor & 2D keypoints & Frame-wise \\
MiliPoint~\cite{cui2023milipoint} & Pose estimation & Radar point cloud & 3D keypoints & Frame-wise \\
mmMesh~\cite{xue2021mmmesh} & Dynamic mesh construction & Radar point cloud & Human mesh & Sequential prior \\
P4Transformer~\cite{fan2021p4transformer} & Point-cloud sequence modeling & Point cloud sequence & Spatio-temporal features & Sequence-level \\
mmBody~\cite{zhang2022mmbody} & Human mesh recovery & Radar point cloud & SMPL mesh & Frame-wise \\
RT-Pose~\cite{ho2024rtpose} & Pose estimation benchmark & Radar tensor & 3D keypoints & Frame-wise \\
RETR~\cite{yataka2024retr} & Indoor radar perception & Radar heatmaps & Detection / segmentation & Frame-wise \\
RT-Mesh~\cite{fan2026m4human} & Human mesh recovery & Radar tensor & SMPL-X mesh & Frame-wise \\
\textbf{RaStream (Ours)} & \textbf{Human mesh recovery} & \textbf{Radar tensor} & \textbf{SMPL-X mesh} & \textbf{Causal streaming} \\
\bottomrule
\end{tabular}
\end{table*}

Table~\ref{tab:related_work_comparison} positions RaStream relative to representative RF/mmWave sensing and reconstruction methods. Point-cloud approaches operate on selected reflections, while radar-tensor or heatmap methods preserve denser spatial evidence but generally perform frame-wise keypoint estimation, detection, segmentation, or mesh recovery. Sequence models provide temporal context, but do not combine native-tensor spatial encoding with persistent, fixed-state causal refinement for streaming SMPL-X recovery. RaStream brings these components together within a single edge-oriented framework.

The main contributions of this paper are:

\begin{itemize}

\item RaStream introduces a localization-conditioned spatial encoder that preserves the native 3D radar structure, separates global localization from subject-centered reconstruction, and fuses global and local evidence into a compact spatial token for temporal reasoning.

\item RaStream performs causal temporal reasoning in compact token space, enabling fixed-memory history aggregation without future observations, long raw-radar histories, or recurrent processing of high-dimensional radar volumes; its slow morphology and fast motion states provide an additional consistent refinement beyond a comparable single-state recurrent baseline.

\item RaStream is evaluated across spatial-backbone scale, streaming stride, and temporal unroll horizon, with reconstruction, temporal consistency, latency, memory, and power profiling on M4Human and Jetson Orin Nano.

\end{itemize}

The remainder of this paper is organized as follows. Section~\ref{sec:overview} provides an overview of RaStream, including the problem formulation and complete architecture. Sections~\ref{sec:spatial_encoding} and \ref{sec:temporal_refinement} detail the spatial encoder and dual-state causal temporal refiner, respectively. Section~\ref{sec:deployment_design} presents the deployment-aware streaming design. Section~\ref{sec:experiments} reports experimental results, ablations, and edge profiling. Section~\ref{sec:conclusion} concludes the paper and discusses remaining limitations.

\section{Overview}
\label{sec:overview}

\subsection{Problem Formulation}
\label{sec:problem_formulation}

We formulate streaming radar mesh recovery as a causal, fixed-memory state
estimation problem. At each emitted step, the model receives only the most
recent radar observation and persistent states propagated from preceding
steps, and must predict the SMPL-X state for the current instant.

Let $X_t \in \mathbb{R}^{T_w \times H \times W \times D}$ denote the short-window mmWave radar tensor at time step $t$, where $T_w$ is the number of consecutive radar frames and $H$, $W$, $D$ are the three radar spatial dimensions. Here, $t$ indexes streaming observations or emitted tokens rather than raw radar frames; consecutive observations are separated by the streaming stride $s$ introduced in Section~\ref{sec:temporal_sampling}.

At streaming step $t$, $X_t$ is formed from the latest $T_w$ radar frames and
the window advances according to the streaming stride $s$. Consecutive inputs
can therefore overlap when $s<T_w$. The radar frames are concatenated along
the input-channel dimension, while the three spatial axes $(H,W,D)$ are
preserved. This defines a short spatio-temporal input for the spatial encoder
without peak extraction, point-cloud conversion, or spatial-axis flattening.
The same windowing rule is used during training and causal inference.

The goal is to estimate a sequence of parametric human body states
\begin{equation}
y_t = \left(\beta_t,\varphi_t,\omega_t,p_t,g_t\right),
\end{equation}
where $\beta_t \in \mathbb{R}^{10}$ denotes the SMPL-X body-shape parameters, $\varphi_t$ denotes the articulated body pose, $\omega_t$ denotes the global root orientation, $p_t \in \mathbb{R}^{3}$ denotes the global translation, and $g_t \in [0,1]$ denotes the predicted gender probability~\cite{pavlakos2019smplx}.

For edge deployment, the prediction at time $t$ must depend on the current and
preceding observations $\{X_{\tau}\}_{\tau \leq t}$ only; it cannot access
future frames. We represent the bounded memory available from preceding steps
by fixed-size states $u_{t-1}$ and $h_{t-1}$, giving the causal recurrence
\begin{equation}
(\hat{y}_t,u_t,h_t)
=F_{\mathrm{RaStream}}(X_t,u_{t-1},h_{t-1}).
\end{equation}
This causal edge-deployment constraint distinguishes our setting from offline
temporal mesh and pose models that can leverage full video
context~\cite{kocabas2020vibe,choi2022mps-net,lin2023motionbert}. It also rules
out retaining long raw-radar histories or recurrently processing
high-dimensional volumetric features. The next subsection introduces the
resulting spatial--temporal decomposition of this recurrence.

\subsection{RaStream Overview}
\label{sec:rastream_overview}

\begin{figure*}[t]
\centering
\includegraphics[width=0.95\textwidth]{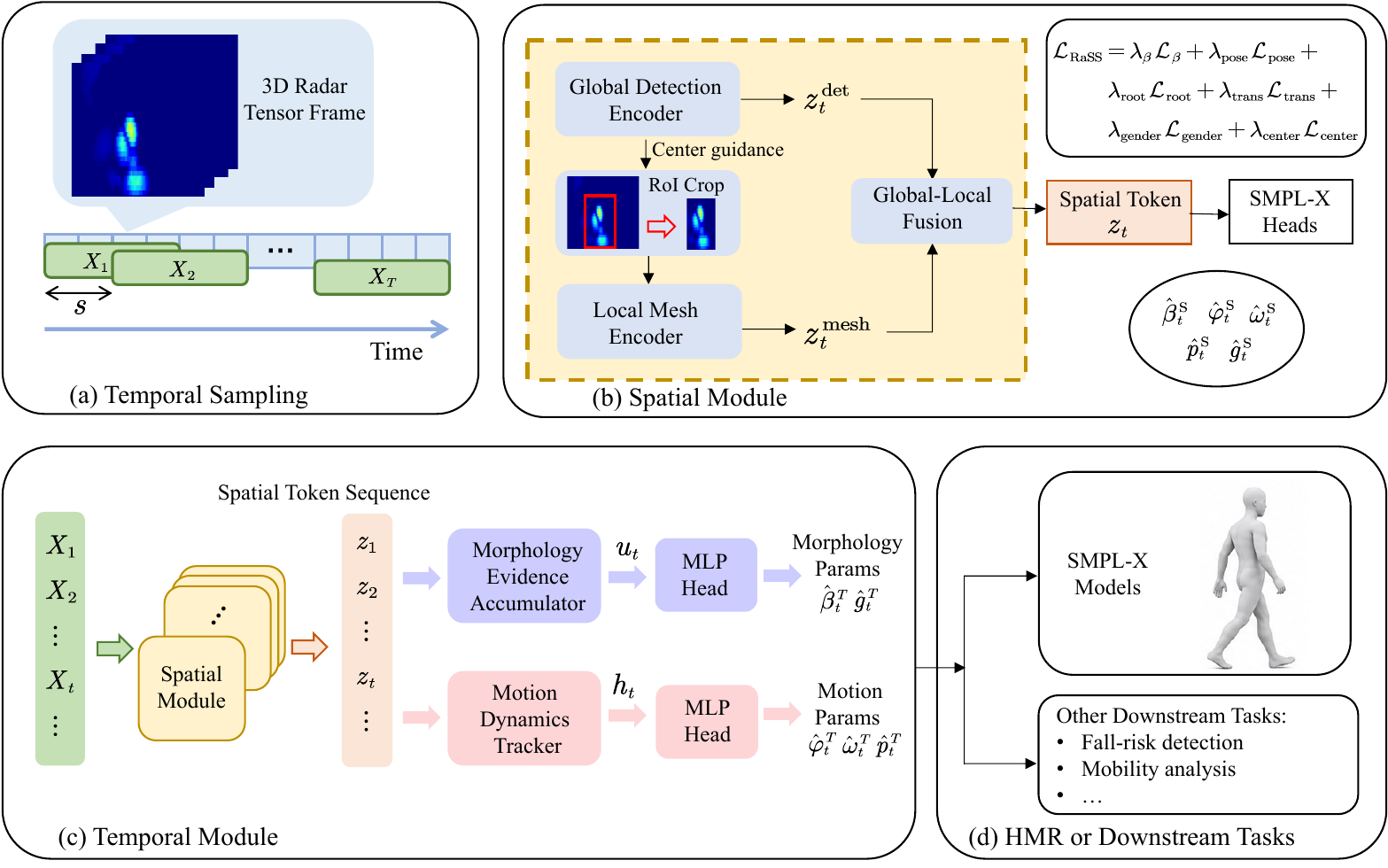}
\caption{Overall architecture of RaStream. Each short radar window passes through spatial encoding, compact-token causal temporal refinement, and SMPL-X parameter regression. The dual-state refiner updates morphology state $u_t$ and motion state $h_t$ without future observations or recurrent processing of raw radar volumes.}
\label{fig:framework}
\end{figure*}

Figure~\ref{fig:framework} summarizes RaStream. Each current radar window is
processed once by a localization-conditioned spatial encoder, which preserves
the volumetric radar structure while combining global scene context with
subject-centered local evidence. The encoder compresses this rich observation
into a compact spatial token,
\begin{equation}
z_t=f_{\mathrm{sp}}(X_t;\theta_{\mathrm{sp}}).
\end{equation}
The token is the interface between costly spatial perception and lightweight
streaming inference: high-dimensional radar processing is confined to the
current window, whereas temporal history is represented only by persistent
compact states.

For single-window evaluation, an auxiliary head produces
\begin{equation}
\hat{y}^{\mathrm{sp}}_t=h_{\mathrm{sp}}(z_t),
\end{equation}
which establishes the body information available from the current observation
alone. The full model refines this estimate causally through
\begin{equation}
(\hat{y}_t,u_t,h_t)=f_{\mathrm{temp}}(z_t,u_{t-1},h_{t-1}).
\end{equation}
The morphology state $u_t$ accumulates slowly varying evidence for body shape
and gender, while the motion state $h_t$ tracks rapidly changing pose, root
orientation, and translation. This division lets RaStream improve temporal
stability without making morphology estimates overly responsive or suppressing
genuine motion.

\section{RaStream Spatial Encoding}
\label{sec:spatial_encoding}

\subsection{Spatial Design Principles}
\label{sec:spatial_principles}

Dense radar-based human mesh recovery requires both full-scene context and spatially concentrated body evidence. The full radar volume is needed to localize the subject, whereas fine-grained articulated geometry depends on weak reflections from limbs and other body parts. Fine-grained multi-user hand sensing has likewise considered target localization before detailed pose estimation~\cite{peng2026milli2hands}, although its hand-pose setting differs from full-body SMPL-X recovery. Applying uniform mesh reasoning to the entire volume spends computation on background regions, but processing only a local crop requires a reliable spatial anchor. RaStream therefore decomposes spatial perception into global body localization followed by subject-centered local reconstruction.

The spatial representation must also retain weak radar evidence and the geometry of the sensing grid. Peak selection and point-cloud conversion can remove low-amplitude responses, while folding one radar axis into the channel dimension weakens the native 3D organization of the tensor. RaStream instead preserves the three radar axes $(H,W,D)$ throughout spatial feature extraction. The $T_w$ frames in each short window are organized only along the input-channel dimension, allowing 3D convolutions to operate on the original radar geometry.

These principles define the interface between spatial perception and temporal reasoning. The full volume provides coarse localization and global context, the center-guided RoI focuses computation on subject-centered evidence, and gated global--local fusion produces the compact spatial token $z_t$. This token supports both single-window prediction and the causal temporal refinement in Section~\ref{sec:temporal_refinement}. For direct spatial supervision and evaluation, the RaStream (spatial-only) mapping is
\begin{equation}
\hat{y}^{\mathrm{sp}}_t = h_{\mathrm{sp}}(f_{\mathrm{sp}}(X_t; \theta_{\mathrm{sp}})),
\end{equation}
where $\theta_{\mathrm{sp}}$ denotes the spatial encoder parameters. The RaStream (spatial-only) output can still exhibit temporal jitter under sparse and noisy observations, motivating the fixed-state temporal processing described in the next section.

\subsection{Spatial Encoder Architecture}
\label{sec:spatial_architecture}

\begin{figure}[t]
\centering
\includegraphics[width=0.95\columnwidth]{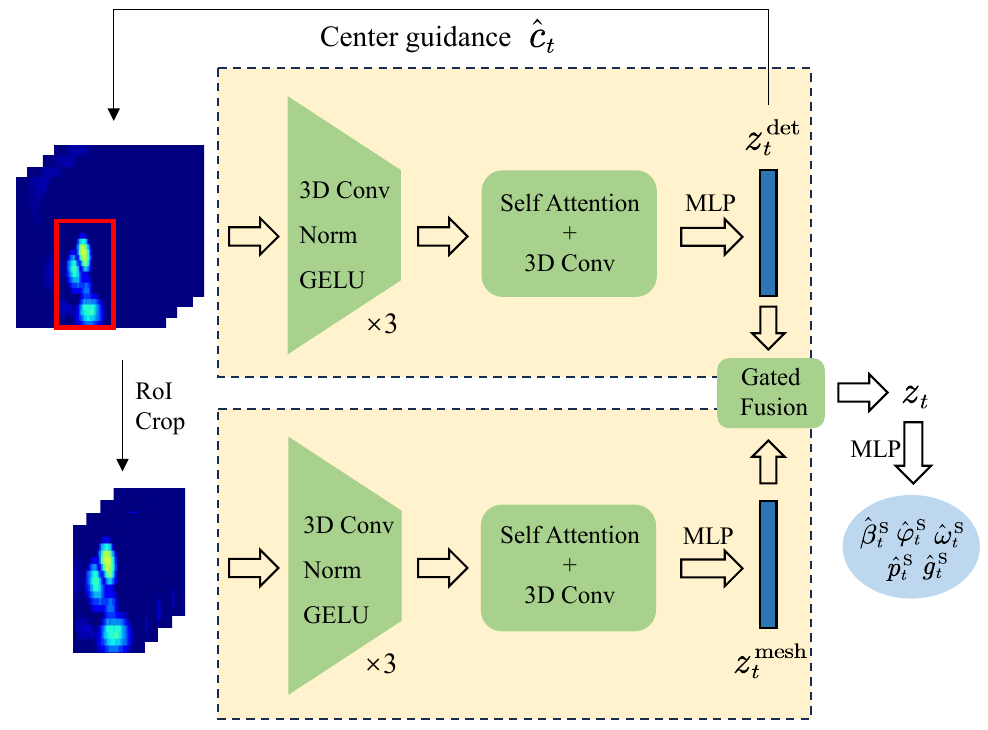}
\caption{RaStream spatial encoder. The localization-conditioned pipeline estimates the body center from the full radar volume, extracts a subject-centered RoI for the local branch, and fuses global and local features before SMPL-X parameter regression.}
\label{fig:spatial_encoder_architecture}
\end{figure}

The RaStream spatial encoder is organized as a two-stage radar-aware model:
\begin{equation}
X_t
\xrightarrow{\;E_{\text{det}}\;}
z^{\text{det}}_t,\hat{c}_t
\xrightarrow{\;\Pi(\cdot,\hat{c}_t)\;}
\tilde{X}_t
\xrightarrow{\;E_{\text{mesh}}\;}
z^{\text{mesh}}_t
\xrightarrow{\;\Phi\;}
\hat{y}^{\mathrm{sp}}_t,
\end{equation}
where $E_{\text{det}}$ is the global detection encoder, $\hat{c}_t$ is the predicted coarse body center, $\Pi(\cdot,\hat{c}_t)$ is the center-guided RoI crop operator, $E_{\text{mesh}}$ is the local mesh encoder, and $\Phi$ denotes the final global-local fusion and regression heads.

The spatial encoder follows the two-stage pipeline in Figure~\ref{fig:spatial_encoder_architecture}. The detection branch encodes the full radar volume and predicts a coarse body center $\hat{c}_t$:
\begin{equation}
z^{\text{det}}_t = E_{\text{det}}(X_t), \quad \hat{c}_t = h_c(z^{\text{det}}_t),
\end{equation}
where $E_{\text{det}}$ is a 3D CNN backbone followed by a single-scale 3D FPN block with residual multi-head spatial self-attention, and $h_c$ is a lightweight MLP head. The encoder uses adaptive pooling to aggregate sparse radar features into a compact descriptor $z^{\text{det}}_t$. Given the predicted center, a differentiable RoI crop operator $\Pi(\cdot, \hat{c}_t)$ extracts a subject-centered local volume $\tilde{X}_t = \Pi(X_t, \hat{c}_t)$. The mesh branch then encodes this crop to obtain a local descriptor:
\begin{equation}
z^{\text{mesh}}_t = E_{\text{mesh}}(\tilde{X}_t).
\end{equation}
The same single-scale FPN-attention block is used in the mesh branch; Table~\ref{tab:model_scale_config} reports the number of attention heads used by the tiny, small, and base variants.

The global descriptor $z^{\text{det}}_t$ and local descriptor $z^{\text{mesh}}_t$ are combined through gated fusion:
\begin{subequations}
\begin{align}
\alpha_t &= \sigma(W_\alpha [z^{\text{det}}_t; z^{\text{mesh}}_t] + b_\alpha), \\
z_t &= \alpha_t \odot z^{\text{det}}_t + (1-\alpha_t) \odot z^{\text{mesh}}_t,
\end{align}
\end{subequations}
where $\sigma$ is the sigmoid function, and $\odot$ denotes element-wise multiplication. The gate $\alpha_t$ adaptively balances global and local information based on observation characteristics. The spatial token $z_t$ is passed to task-specific MLP heads that predict shape $\hat{\beta}$, body pose $\hat{\varphi}$, root orientation $\hat{\omega}$, translation $\hat{p}$, and gender $\hat{g}$, following parametric mesh recovery~\cite{kanazawa2018hmr,kolotouros2019spin,kocabas2020vibe} with radar-aware tokens. Pose and root orientation are regressed with the continuous 6D rotation representation~\cite{zhou2019continuity} rather than axis-angle targets, which avoids discontinuities that can amplify ambiguity from weak or missing radar reflections. The network predicts the first two columns of a rotation matrix and recovers a valid matrix by Gram-Schmidt orthonormalization, denoted as $\tilde{R}=\operatorname{GS}(v)$; this conversion is applied before SMPL-X decoding, and pose loss is computed on recovered rotation matrices rather than raw 6D vectors. Together with $\hat{c}_t$, these form the RaStream (spatial-only) output $\hat{y}^{\mathrm{sp}}_t$.

\subsection{Spatial Training Objective}
\label{sec:spatial_objective}
The single-window objective is formulated as a weighted multi-task loss
\begin{align}
\mathcal{L}_{\text{sp}}
=\;& \lambda_{\beta}\mathcal{L}_{\beta} +
\lambda_{\text{pose}}\mathcal{L}_{\text{pose}} +
\lambda_{\text{root}}\mathcal{L}_{\text{root}} \nonumber\\
&+ \lambda_{\text{trans}}\mathcal{L}_{\text{trans}} +
\lambda_{\text{gender}}\mathcal{L}_{\text{gender}} +
\lambda_{\text{center}}\mathcal{L}_{\text{center}}.
\end{align}
Here, $\mathcal{L}_{\beta}$, $\mathcal{L}_{\text{pose}}$, $\mathcal{L}_{\text{root}}$, $\mathcal{L}_{\text{trans}}$, $\mathcal{L}_{\text{gender}}$, and $\mathcal{L}_{\text{center}}$ supervise shape, body pose, root orientation, translation, gender, and body center, respectively. The scalar coefficients balance the relative importance of the six tasks; the concrete loss forms and weights are reported with the implementation details in Section~\ref{sec:training_protocol}.
The center term directly supervises the anchor used for RoI cropping. Localization errors can shift the crop toward background or exclude weak limb reflections, degrading mesh and translation estimates; joint supervision therefore couples localization and reconstruction.

\section{Dual-State Causal Temporal Refinement}
\label{sec:temporal_refinement}

\subsection{Temporal Design Principles}
\label{sec:temporal_principles}

Frame-local radar ambiguity can destabilize mesh estimates, motivating causal temporal refinement. Rather than buffering or recurrently processing radar volumes, RaStream processes each window once and propagates history only through compact token $z_t$ and fixed-size states. Because morphology varies slowly while pose and translation change rapidly, RaStream separates temporal state into a morphology accumulator $u_t$ and motion tracker $h_t$. This dual-state design refines generic causal recurrence while preserving responsiveness to current observations.

\subsection{Dual-State Latent Refinement}
\label{sec:dual_state_refinement}

The dual-state causal temporal refiner receives the spatial token from the pretrained localization-conditioned spatial encoder. Let
\begin{equation}
z_t = f_{\mathrm{sp}}(X_t)
\end{equation}
denote the token extracted from the fused spatial feature stream, where $f_{\mathrm{sp}}(\cdot)$ summarizes the single-window encoder and pooling pipeline up to the shared representation used by the output heads. The temporal mapping is
\begin{equation}
f_{\text{temp}}: (z_t,u_{t-1},h_{t-1}) \mapsto (\hat{y}_t,u_t,h_t),
\end{equation}
where $u_t$ is the morphology state, $h_t$ is the motion state, and $\hat{y}_t$ is the SMPL-X state predicted by the complete RaStream framework.

\begin{algorithm}[H]
\caption{Causal streaming inference in RaStream}
\label{alg:streaming_inference}
\begin{minipage}{0.94\columnwidth}
\setcounter{algline}{0}
\small
\textbf{Input:} radar stream, window length $T_w$, stride $s$\\
\textbf{Output:} emitted SMPL-X meshes\\[-2pt]
\algline{$u_0\gets\mathbf{0}$; $h_0\gets\mathbf{0}$}
\algline{\textbf{for} each emitted step $t$ \textbf{do}}
\algline{\hspace{0.18in}Form $X_t$ from the latest radar frames}
\algline{\hspace{0.18in}$z_t\gets f_{\mathrm{sp}}(X_t)$}
\algline{\hspace{0.18in}$s_t\gets E_{\mathrm{static}}(z_t)$; $a_t\gets\sigma(g_{\mathrm{static}}(z_t))$}
\algline{\hspace{0.18in}$u_t\gets(1-a_t)u_{t-1}+a_ts_t$}
\algline{\hspace{0.18in}$m_t\gets E_{\mathrm{dynamic}}(z_t)$}
\algline{\hspace{0.18in}$h_t\gets F_{\mathrm{dynamic}}(h_{t-1},m_t)+R_{\mathrm{dynamic}}(m_t)$}
\algline{\hspace{0.18in}$(\hat{\beta}_t,\hat{g}_t)\gets H_{\mathrm{static}}(u_t)$}
\algline{\hspace{0.18in}$(\hat{\varphi}_t,\hat{\omega}_t,\hat{p}_t)\gets H_{\mathrm{dynamic}}(h_t,u_t)$}
\algline{\hspace{0.18in}Decode the parameters locally or transmit them for cloud decoding}
\algline{\textbf{end for}}
\end{minipage}
\end{algorithm}

The streaming states and SMPL-X parameters are maintained at the edge. If a
full mesh is not required locally, the predicted parameters can be transmitted
to a remote server for mesh decoding and visualization. The temporal refiner operates on the 512--1024 dimensional spatial token $z_t$. The morphology and motion encoders are two-layer MLPs with layer normalization, GELU activation, and dropout.

The morphology branch extracts a single-frame morphology observation
\begin{equation}
s_t = E_{\text{static}}(z_t),
\end{equation}
which encodes stable body scale, limb-proportion, and subject-identity cues. The morphology evidence accumulator maintains the slow state through a token-conditioned exponential moving average (EMA):
\begin{equation}
u_t = (1-a_t)u_{t-1}+a_t s_t,\qquad a_t\in[0,1],
\label{eq:morph_state}
\end{equation}
where $a_t$ is an adaptive EMA update coefficient. The learned gate allows the model to assign different update weights to different radar observations, reducing the influence of frames that are less useful for stable morphology estimation. A morphology prediction head maps this accumulated representation to morphology-related SMPL-X attributes:
\begin{equation}
(\hat{\beta}_t,\hat{g}_t) = H_{\text{static}}(u_t).
\end{equation}
The state aggregates morphology evidence across the stream instead of estimating these attributes independently at every step.
The motion branch extracts the fast-varying feature
\begin{equation}
m_t = E_{\text{dynamic}}(z_t),
\end{equation}
and updates a causal recurrent state:
\begin{equation}
h_t = F_{\text{dynamic}}(h_{t-1},m_t)+R_{\text{dynamic}}(m_t).
\end{equation}
In our implementation, $F_{\text{dynamic}}$ is a unidirectional GRU~\cite{cho2014rnnencoderdecoder,chung2014gru} and $R_{\text{dynamic}}$ is a lightweight projection residual from the current motion feature. Since $h_t$ already contains the current motion evidence through this residual recurrent update, the motion prediction head is conditioned on the fast motion state and the slow morphology state:
\begin{equation}
(\hat{\varphi}_t,\hat{\omega}_t,\hat{p}_t)
=
H_{\text{dynamic}}(h_t,u_t).
\end{equation}
This conditioning is necessary because the same joint rotations produce different joints and vertices when applied to different body shapes. The adaptive update weight is regressed directly from the current spatial token:
\begin{equation}
a_t = \sigma(g_{\text{static}}(z_t)),
\end{equation}
where $g_{\text{static}}$ is a small MLP producing a scalar gate. The learned gate modulates how much each token updates the morphology state instead of treating all radar observations equally. The residual recurrent update of $h_t$ keeps motion refinement causal while retaining the contribution of the current token.

\begin{figure}[t]
\centering
\includegraphics[width=\columnwidth]{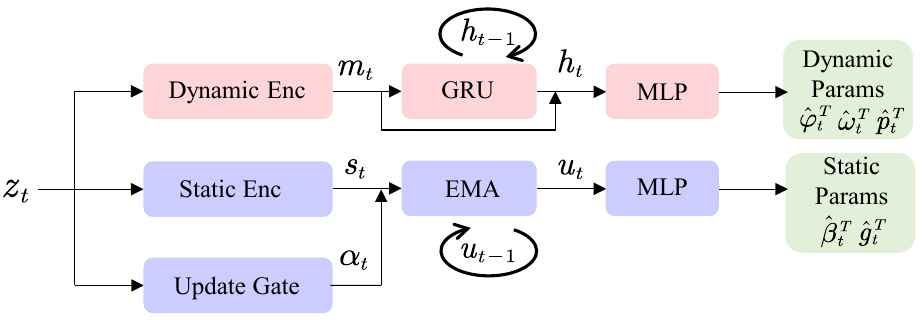}
\caption{Dual-state causal temporal refinement in RaStream. The spatial token $z_t$ is separated into morphology and motion features. The morphology evidence accumulator updates slow state $u_t$ for shape and gender estimation, while the motion dynamics tracker updates fast state $h_t$ for pose, orientation, and translation.}
\label{fig:rastream_temporal_unroll}
\end{figure}

\subsection{Initialization and State Recovery}
\label{sec:state_recovery}

At the beginning of an independent sequence, RaStream initializes both
temporal states to zero. The first valid radar window is processed by the spatial encoder,
after which the states are updated and predictions are emitted at the chosen
stride. The states are reset when a new sequence or subject is encountered.
If observations are temporarily unavailable, the most recent states are
retained. When the stream resumes, cached spatial tokens can be replayed to
re-warm the temporal states without recomputing the corresponding volumetric
windows. This recovery procedure keeps spatial recomputation separate from
temporal state recovery.

\subsection{Temporal Training Objective}
\label{sec:temporal_objective}

The temporal training objective extends the spatial mesh-parameter loss with all-frame supervision, temporal consistency, and morphology-stability terms. Temporal motion losses reduce jitter, following video pose and mesh recovery~\cite{pavllo20193d,kanazawa2019learning,choi2022mps-net}:
\begin{align}
\mathcal{L}_{\text{temp}}
=\;&
\alpha_{\text{tv}}\mathcal{L}_{\text{trans-vel}}
+ \alpha_{\text{pv}}\mathcal{L}_{\text{pose-vel}}
+ \alpha_{\text{rv}}\mathcal{L}_{\text{root-vel}} \nonumber\\
&+
\alpha_{\text{ta}}\mathcal{L}_{\text{trans-acc}}.
\end{align}
The velocity terms compare adjacent prediction and target differences:
\begin{subequations}
\begin{align}
\mathcal{L}_{\text{trans-vel}}
&= \sum_{\tau}\|(\hat{p}_{\tau}-\hat{p}_{\tau-1})
-(p_{\tau}-p_{\tau-1})\|_2^2,\\
\mathcal{L}_{\text{pose-vel}}
&= \sum_{\tau}\|(\hat{\varphi}_{\tau}-\hat{\varphi}_{\tau-1})
-(\varphi_{\tau}-\varphi_{\tau-1})\|_2^2,\\
\mathcal{L}_{\text{root-vel}}
&= \sum_{\tau}\|(\hat{\omega}_{\tau}-\hat{\omega}_{\tau-1})
-(\omega_{\tau}-\omega_{\tau-1})\|_2^2.
\end{align}
\end{subequations}
For pose and root orientation, the differences are computed after converting the 6D representation to rotation matrices, so the loss is applied in matrix space rather than by subtracting axis-angle parameters.
The acceleration term penalizes second-order motion mismatch:
\begin{equation}
\mathcal{L}_{\text{trans-acc}}
= \sum_{\tau}\| \Delta^2\hat{p}_{\tau}-\Delta^2 p_{\tau}\|_2^2,
\end{equation}
where $\Delta^2 p_{\tau}=p_{\tau}-2p_{\tau-1}+p_{\tau-2}$.

The all-frame parameter term supervises every emitted prediction in the finite unroll:
\begin{equation}
\mathcal{L}_{\text{all}}
=
\sum_{\tau=1}^{T}\ell_{\text{param}}(\hat{y}_{\tau},y_{\tau}),
\end{equation}
where $\ell_{\text{param}}$ uses the same shape, pose, root, translation, and gender components as the single-window objective. Within-sequence morphology stability is further encouraged by prediction- and state-level smoothness:
\begin{equation}
\mathcal{L}_{\text{shape-cons}}
=
\sum_{t=2}^{T}\|\hat{\beta}_t-\hat{\beta}_{t-1}\|_2^2.
\end{equation}
\begin{equation}
\mathcal{L}_{\text{state-smooth}}
=
\sum_{t=2}^{T}\|u_t-u_{t-1}\|_2^2.
\end{equation}
These terms encode the prior that body shape is stable within a stream without adding inference-time cost. Let $\mathcal{L}_{\text{last}}=\ell_{\text{param}}(\hat{y}_T,y_T)$ denote the last-frame parameter loss.

The full objective is
\begin{align}
\mathcal{L}
=\;&
\mathcal{L}_{\text{last}}
+ \lambda_{\text{all}}\mathcal{L}_{\text{all}}
+ \lambda_{\text{temp}}\mathcal{L}_{\text{temp}}
+ \lambda_{\text{shape-cons}}\mathcal{L}_{\text{shape-cons}} \nonumber\\
&+
\lambda_{\text{state}}\mathcal{L}_{\text{state-smooth}}.
\end{align}

All emitted steps are supervised rather than only the final prediction, while the velocity and acceleration terms penalize short-term temporal artifacts.

\subsection{Temporal Sampling Parameters}
\label{sec:temporal_sampling}

RaStream uses three temporal sampling parameters: local window length $T_w$, streaming stride $s$, and finite unroll horizon $T$. Unless otherwise stated, $T_w=4$. The effective raw-frame coverage of a finite training or evaluation segment is
\begin{equation}
T_{\text{cov}} = T_w + (T-1)s.
\end{equation}

Here $T_w$ controls local radar evidence, $s$ controls output rate, and $T$ controls the finite training unroll and warm-up/replay horizon. The recurrent states persist during steady-state inference and are not reset after every $T$ tokens. Thus, $T_{\text{cov}}$ describes the physical span of the finite segment, not a limit on recurrent memory.

For example, $T_w=4$, $T=16$, and $s=2$ cover 34 raw radar frames, or approximately 2.83 s at 12 Hz. Increasing $T_w$ increases the cost of each spatial-encoder invocation, increasing $T$ extends the supervised temporal span, and increasing $s$ lowers the output rate while reducing the amortized invocation frequency. We evaluate the effects of $T$ and $s$ in Section~\ref{sec:experiments}, while fixing $T_w=4$ in the reported experiments.

\section{Deployment-Aware Streaming Design}
\label{sec:deployment_design}

\subsection{Scale Variants}
\label{sec:scale_variants}
For deployment, RaStream provides Tiny, Small, and Base spatial-backbone scales as the model-scale variable $m$. Table~\ref{tab:model_scale_config} summarizes the proportional channel scaling; backbone channels are listed as Stage1/Stage2/Stage3.

\begin{table}[t]
\caption{RaStream spatial encoder scale configurations.}
\label{tab:model_scale_config}
\centering
\footnotesize
\setlength{\tabcolsep}{2.4pt}
\begin{tabular}{lccccc}
\toprule
Scale & Backbone & Embedding & MHSA & Params & GFLOPs \\
      & Channels & Dim & Heads & (M) & /window \\
\midrule
Tiny  & 32/64/128  & 512  & 4 & 5.6  & 1.0 \\
Small & 48/96/192  & 768  & 6 & 12.5 & 2.0 \\
Base  & 64/128/256 & 1024 & 8 & 22.2 & 3.2 \\
\bottomrule
\end{tabular}
\end{table}

Backbone channels scale by 0.5$\times$ (tiny), 0.75$\times$ (small), or 1.0$\times$ (base), with downstream dimensions scaled proportionally. The variants span a 4$\times$ parameter range and a 3$\times$ compute range without architectural redesign. The base model targets the highest reconstruction quality, the small model offers an accuracy-efficiency compromise, and the tiny model supports latency-sensitive or low-power deployments.

\subsection{Hardware and Inference Pipeline}
\label{sec:hardware_pipeline}

We deploy RaStream on the NVIDIA Jetson Orin Nano, a representative edge computing platform for embedded sensing and robotics applications. The device features an NVIDIA Ampere GPU with 1024 CUDA cores and Tensor Cores, 8 GB of unified memory shared between CPU and GPU, and typical power consumption of 7-15W under load. The software stack consists of PyTorch with TensorRT optimization for inference acceleration.

The deployment pipeline has two computational stages within RaStream: spatial encoding and dual-state temporal refinement. The spatial encoder produces the compact token and accounts for most inference time because it processes 3D radar volumes. The temporal refiner updates persistent morphology and motion states, applies the token-conditioned gate, and runs SMPL-X heads. Since the states persist across frames, RaStream avoids replaying the finite unroll horizon from scratch at every output.

The system stores a small ring buffer of compact tokens rather than raw radar volumes. Streaming updates $u_t$ and $h_t$ with $O(1)$ temporal cost, minimizing memory traffic on embedded devices~\cite{chen2019deep,deng2020model,lane2016deepx}. If tracking is interrupted or frames are dropped, states can be re-warmed by replaying cached tokens without rerunning the spatial encoder; this replay is lightweight enough for the short warm-up/replay horizons used in our system.

Profiling uses batch size one. Each run executes 10 warm-up inferences before measurement, excludes the first measured frame from steady-state statistics, synchronizes CUDA around timed GPU stages, and reports mean and 95th-percentile end-to-end latency over the remaining stream. Power is read from Jetson VDD\_IN before and after the inference run and reported as idle/active power.

\subsection{Computation and State Memory}
\label{sec:computation_memory}

Let $d_z$ denote the spatial-token dimension and $d_h$ the dimension of the
motion state. The spatial encoder is evaluated once for each emitted token and accounts for
most of the per-invocation computation. The temporal refiner processes only
$z_t$ and updates fixed-dimensional states, so its recurrent computation and
state memory do not grow with stream length. The persistent states require
$O(d_z+d_h)$ memory. If a replay buffer of $K$ tokens is retained, its memory
is $O(Kd_z)$, compared with $O(KT_wHWD)$ for storing the corresponding radar
windows.

The stride changes the invocation frequency but not the cost of an individual
spatial-encoder call. If $C_{\mathrm{sp}}(m)$ and $C_{\mathrm{temp}}(m)$ denote the
costs of the spatial and temporal branches at model scale $m$, the
approximate cost per raw radar frame is
\begin{equation}
C_{\mathrm{frame}}(m,s)
\approx \frac{C_{\mathrm{sp}}(m)+C_{\mathrm{temp}}(m)}{s}.
\end{equation}
This expression is used only to interpret the measured profiling results;
preprocessing and host--device transfer costs are included in the reported
end-to-end measurements.

\subsection{Configuration Space}
\label{sec:configuration_space}

Edge devices have heterogeneous latency, power, memory, and output-rate requirements. Because radar mesh recovery uses sparse 4D volumetric tensors, its accuracy-efficiency tradeoffs differ from RGB-based systems. We therefore make the main deployment variables explicit.

A deployment configuration is $x=(m,T_w,T,s,q)$, where $m$ is model scale, $T_w$ is local radar window length, $T$ is the finite unroll and warm-up/replay horizon, $s$ is streaming stride, and $q$ is inference precision. The experiments fix $T_w=4$ and $q=\mathrm{FP32}$, and vary $m$, $T$, and $s$. The RaStream (spatial-only) variant is used as the non-temporal reference.

The framework-level deployment configuration space is
\begin{equation}
\Omega=\mathcal{M}\times\mathcal{T}_w\times\mathcal{T}\times\mathcal{S}\times\mathcal{Q},
\end{equation}
where $\mathcal{M}=\{\text{Tiny},\text{Small},\text{Base}\}$, $\mathcal{T}_w=\{2,4,8\}$, $\mathcal{T}=\{4,8,16\}$, $\mathcal{S}=\{1,2,4\}$, and $\mathcal{Q}$ denotes the set of precision modes supported by the target hardware.

The actually evaluated subset is
\begin{equation}
\Omega_{\mathrm{eval}}=\mathcal{M}\times\{T_w=4\}\times\mathcal{T}\times\mathcal{S}\times\{\mathrm{FP32}\},
\end{equation}
Only this subset is enumerated in the experiments and profiling; $T_w$ and $q$ remain configurable framework dimensions rather than systematically evaluated dimensions in this paper. For a target application, the measured configurations can be ranked by mesh error, temporal quality (TA and JA), and energy per output, then filtered to satisfy P95 latency, memory, power, output-rate, and steady-state end-to-end latency requirements. Energy per output normalizes average power by the output rate.

Steady-state end-to-end latency is decomposed as:
\begin{equation}
\begin{aligned}
D_{\text{steady}}(x)
=
L_{\text{sp}}(m,T_w,q)
+
L_{\text{state}}(m,q)
\\
{}+
L_{\text{head}}(m,q)
+
L_{\text{post}}
+
L_{\text{sync}},
\end{aligned}
\end{equation}
where the terms denote spatial encoding, state update, prediction heads, SMPL-X decoding, and host-device synchronization.

The measured candidate set can be enumerated and pruned for Pareto dominance. Section~\ref{sec:experiments} reports the resulting FP32 Jetson profiling and Pareto analysis.

\section{Experiments}
\label{sec:experiments}

\subsection{Dataset and Evaluation Protocol}
\label{sec:dataset_protocol}

\textit{Dataset and protocol.}
We evaluate on M4Human~\cite{fan2026m4human}, a large-scale mmWave radar
human mesh benchmark comprising 661K synchronized radar--mesh frames, 50
actions, and 20 subjects. The synchronized observations provide paired radar
tensors and SMPL-X supervision across static and dynamic body motions, making
the benchmark suitable for evaluating both single-window reconstruction and
causal temporal refinement. Following the benchmark protocol, we use a
75\%/5\%/20\% train/validation/test split, corresponding to approximately
496K/34K/132K synchronized samples. Our reported results use the S1 random
split, in which clips are assigned randomly to non-overlapping subsets, and
are grouped into In-Place (IP, 30 actions), Sit-In-Place (SIP, 5 actions),
Non-In-Place (NIP, the remaining non-in-place actions), and ALL. This grouping
follows the benchmark protocols and separates local daily/rehabilitation
motions, seated motions, and larger-displacement activities. Cross-subject,
cross-action, and cross-device generalization are not
evaluated in this study.

\textit{Spatial metrics.}
We report the four M4Human spatial metrics in the world coordinate frame,
without root alignment or Procrustes alignment; lower values indicate better
reconstruction. Let $V=10{,}475$ be the number of SMPL-X vertices,
$J=22$ the number of evaluated joints, $\hat{v}_{i}$ and $v_i$ the predicted
and ground-truth vertices, $\hat{q}_{j}$ and $q_j$ the corresponding joint
positions, $\hat{R}_{j}$ and $R_j$ the predicted and ground-truth joint
rotations, and $\hat{p}$ and $p$ the predicted and ground-truth global root
translations. Mean Vertex Error (MVE) is
\begin{equation}
\mathrm{MVE}=\frac{1}{V}\sum_{i=1}^{V}\|\hat{v}_{i}-v_i\|_2.
\end{equation}
Mean Joint Error (MJE) is
\begin{equation}
\mathrm{MJE}=\frac{1}{J}\sum_{j=1}^{J}\|\hat{q}_{j}-q_j\|_2.
\end{equation}
Mean Rotation Error (MRE) measures the geodesic angular error on $SO(3)$:
\begin{equation}
\mathrm{MRE}=\frac{1}{J}\sum_{j=1}^{J}
\arccos\!\left(\frac{\operatorname{tr}(\hat{R}_{j}R_j^{\mathsf T})-1}{2}\right),
\end{equation}
reported in degrees. Translation Error (TE) is the global root-translation
error
\begin{equation}
\mathrm{TE}=\|\hat{p}-p\|_2.
\end{equation}
MVE, MJE, and TE are reported in millimeters.

\textit{Temporal metrics.}
To evaluate temporal consistency beyond single-frame accuracy, Translation Acceleration (TA) measures global translation smoothness:
\begin{equation}
\mathrm{TA} = \frac{1}{\Delta t^2}\mathbb{E}_{t}\left[\left\|p_{t+1}-2p_t+p_{t-1}\right\|_2\right],
\end{equation}
where $p_t$ is the predicted translation and $\Delta t$ is the physical interval. Joint Acceleration (JA) applies the same second-order measure to per-joint trajectories:
\begin{equation}
\mathrm{JA} = \frac{1}{\Delta t^2}\mathbb{E}_{t}\left[\frac{1}{J}\sum_{j=1}^{J}\left\|q_{t+1}^{j}-2q_t^{j}+q_{t-1}^{j}\right\|_2\right],
\end{equation}
where $q_t^j$ is joint $j$ at frame $t$.

Both metrics use $\Delta t = s / f_{\mathrm{radar}}$, where $f_{\mathrm{radar}} = 12$ Hz and $s$ is the stride. This converts finite differences into physical acceleration units, but it does not remove the effect of temporal subsampling: larger strides change the sampled trajectory and may suppress high-frequency jitter before acceleration is computed. We therefore interpret TA and JA with stride-matched references rather than comparing raw values across different strides.
These metrics are complementary to MVE and MJE. MVE and MJE average independent frame errors, so two methods can obtain similar spatial accuracy while producing very different motion quality. TA and JA instead measure second-order global translation changes and second-order joint trajectory changes. This distinction is important for radar sensing because frame-local ambiguities can create short high-frequency artifacts: a missing limb reflection may not dominate average reconstruction error, but it can disturb activity recognition, gait analysis, rehabilitation monitoring, or human-robot interaction. We therefore report temporal smoothness alongside spatial accuracy rather than treating it as only a qualitative property.

\textit{Deployment metrics.}
Deployment metrics include mean latency, P95 latency, memory usage, and power consumption. Unless otherwise stated, profiling uses FP32 inference on Jetson Orin Nano and reports the measured end-to-end path for the evaluated configuration rather than an operator-by-operator latency decomposition.

\subsection{Training and Inference Protocol}
\label{sec:training_protocol}

The RaStream spatial encoder is trained first with the single-window objective
described in Section~\ref{sec:spatial_encoding}, using synchronized radar--mesh pairs. After training,
the selected spatial checkpoint is applied to the radar stream to produce the
compact tokens used by the causal temporal refiner. The refiner is trained on finite token segments with
the temporal objective in Section~\ref{sec:temporal_refinement}. The recurrent states are initialized at
the beginning of each sampled segment during training. At inference time, the
states are carried from one emitted prediction to the next and reset when a
new sequence starts; no future radar frames are used. The stride determines
when a prediction is emitted. During steady-state inference, $u_t$ and $h_t$
carry the temporal context; cached tokens are retained only for warm-up or
state recovery after an interruption.

\textit{Training details.}
Implementation details follow the configuration used for the reported runs. The spatial encoder uses batch size 64, learning rate $2\times10^{-4}$, three epochs, and step decay with factor 0.8. Its loss weights are 0.3 for shape, 15 for body pose, 1 for root orientation, 10 for translation and center anchoring, and 0.5 for gender; vertex and joint losses are disabled during training and used only for evaluation. The causal temporal refiner uses cached spatial tokens, batch size 256, 20 epochs, AdamW with learning rate $3\times10^{-4}$ and weight decay $10^{-4}$, a two-layer GRU motion state with hidden dimension 512, and zero-initialized recurrent states at the start of each sampled segment. The temporal loss weights are 0.3 for all-frame supervision, 0.5 for translation velocity, 0.1 for pose velocity, 0.1 for root-orientation velocity, 0.1 for translation acceleration, 0.1 for shape consistency, and 0.01 for shape-state smoothness. The best checkpoint is selected by validation loss.

\textit{Baseline protocol.}
The external baseline numbers in Table~\ref{tab:spatial_main_results} follow the M4Human mesh-recovery benchmark protocol. Methods whose original papers target keypoints, detection, segmentation, or point-cloud sequence representation are evaluated through the benchmark's adapted SMPL/SMPL-X regression setting, so MVE, MJE, MRE, and TE are computed from reconstructed parametric meshes rather than from their original task outputs alone. Since existing radar-tensor SMPL-X baselines are predominantly frame-wise, we separate the evaluation into two stages: the RaStream (spatial-only) variants are compared with prior methods under the common single-window protocol, and the complete framework is evaluated through controlled causal refinements using the same pretrained spatial backbone. The dual-state design is compared with a single-state recurrent baseline.

\subsection{Single-Window Spatial Results}
\label{sec:spatial_results}
Table~\ref{tab:spatial_main_results} compares the RaStream (spatial-only) variants with single-window baselines across action groups. Params are in millions; grouped cells report In-Place / Sit-In-Place / Non-In-Place, with ALL shown separately.
All three RaStream (spatial-only) scales improve over RT-Mesh in the ALL setting with fewer parameters. RaStream (spatial-only, Base) reduces ALL MVE from 90.90 mm to 84.27 mm with 22.2M parameters versus RT-Mesh's 63.3M, while the Tiny and Small variants reach 87.5 mm and 85.8 mm. The largest gain appears on Sit-In-Place actions, where RaStream (spatial-only, Base) reduces MVE from 118.10 mm to 63.15 mm. The grouped results show that performance varies with action type, so the aggregate score alone does not capture the full behavior of the models.

\begin{table*}[t]
\caption{Single-window reconstruction results across action groups.}
\label{tab:spatial_main_results}
\centering
\footnotesize
\setlength{\tabcolsep}{1.0pt}
\begin{tabular*}{\textwidth}{@{\extracolsep{\fill}}lccccccccc@{}}
\toprule
\multirow{2}{*}{Method} & \multirow{2}{*}{Params} & \multicolumn{4}{c}{IP / SIP / NIP} & \multicolumn{4}{c}{ALL} \\
\cmidrule(lr){3-6} \cmidrule(lr){7-10}
& (M) & MVE & MJE & MRE & TE & MVE & MJE & MRE & TE \\
\midrule
mmMesh~\cite{xue2021mmmesh} & 41.5 & 105.9/183.3/201.4 & 88.0/164.2/170.6 & 11.0/10.7/15.4 & 57.2/92.8/105.4 & 132.7 & 112.1 & 11.8 & 70.4 \\
P4Transformer~\cite{fan2021p4transformer} & 129.0 & 71.5/115.2/139.7 & 59.0/109.3/121.7 & 8.0/9.5/13.5 & 37.3/69.1/77.9 & 89.5 & 76.6 & 9.2 & 48.6 \\
RT-Pose~\cite{ho2024rtpose} & 6.1 & 80.0/130.2/158.5 & 66.5/126.0/139.3 & 8.7/9.9/14.2 & 43.0/81.6/92.2 & 100.7 & 87.0 & 9.9 & 56.7 \\
RETR~\cite{yataka2024retr} & 53.1 & 73.2/133.6/162.6 & 58.8/126.2/140.4 & 7.5/10.0/14.2 & 34.8/80.4/90.1 & 97.1 & 81.8 & 9.1 & 50.4 \\
RT-Mesh~\cite{fan2026m4human} & 63.3 & 72.4/118.1/142.0 & 59.1/112.2/123.1 & 8.3/9.7/13.8 & 36.2/68.4/77.1 & 90.9 & 77.2 & 9.6 & 47.6 \\
\midrule
RaStream (spatial-only, Tiny) & 5.6 & 70.5/66.8/127.5 & 57.2/55.9/112.8 & 8.1/7.8/12.5 & 35.0/36.2/74.6 & 87.5 & 73.8 & 9.3 & 46.5 \\
RaStream (spatial-only, Small) & 12.5 & 69.3/64.7/125.0 & 56.1/54.0/110.2 & 7.9/7.5/12.2 & 34.3/35.0/73.0 & 85.8 & 72.3 & 9.1 & 45.9 \\
RaStream (spatial-only, Base) & 22.2 & \textbf{68.6/63.2/122.6} & \textbf{55.6/52.7/108.5} & \textbf{7.8/7.4/12.0} & \textbf{34.0/34.2/71.9} & \textbf{84.3} & \textbf{71.2} & \textbf{9.0} & \textbf{45.4} \\
\bottomrule
\end{tabular*}
\end{table*}

\begin{figure}[t]
\centering
\includegraphics[width=\columnwidth]{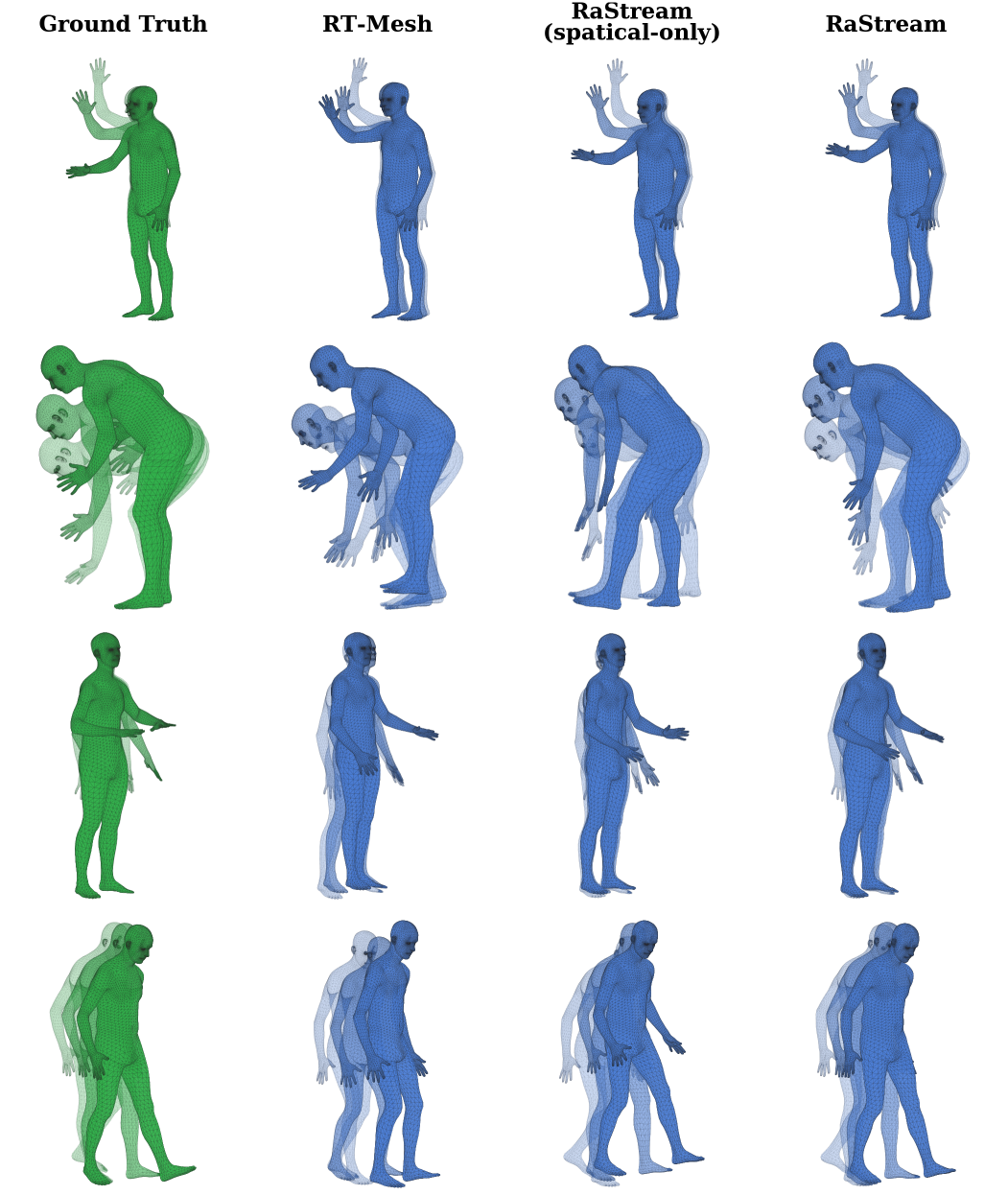}
\caption{Qualitative mesh reconstruction comparison across four representative test sequences. Columns from left to right: Ground Truth, RT-Mesh, RaStream (spatial-only, Base), and RaStream. The snapshots illustrate improved body structure and limb articulation.}
\label{fig:mesh_comparison_4x4}
\end{figure}

Figure~\ref{fig:mesh_comparison_4x4} qualitatively compares ground truth, RT-Mesh, the RaStream (spatial-only, Base) variant, and the complete RaStream framework across four test samples. The snapshots illustrate improved spatial reconstruction fidelity, especially around limb articulations and body contours.
Compared with RT-Mesh, the RaStream (spatial-only) variant recovers cleaner body extent and more plausible limbs, which is consistent with the spatial gains in Table~\ref{tab:spatial_main_results}. Temporal behavior is analyzed separately using quantitative smoothness metrics and trajectory curves in the following sections.

\subsection{Causal Temporal Results}
\label{sec:temporal_results}

\begin{figure*}[!t]
\centering
\includegraphics[width=\textwidth]{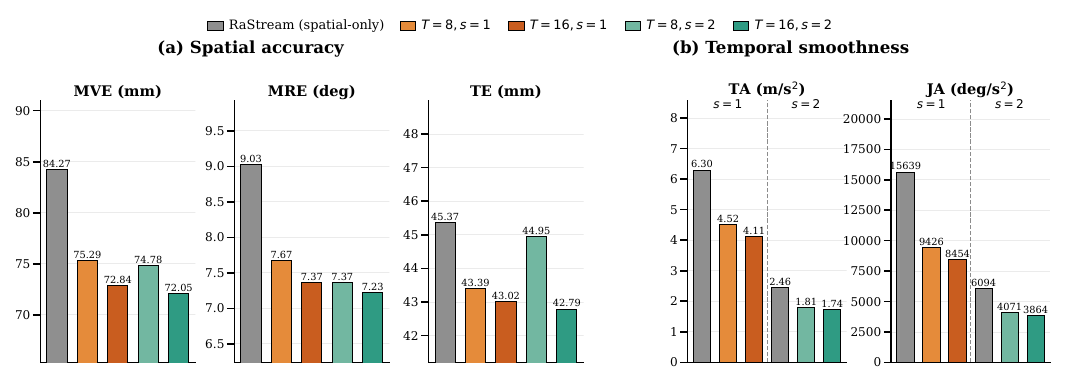}
\caption{RaStream results with different temporal configurations on the ALL action group. (a) Spatial accuracy is shown using raw MVE, MRE, and TE values. (b) Temporal smoothness is shown using raw TA and JA values with stride-matched RaStream (spatial-only) references. Lower is better for all metrics.}
\label{fig:rastream_results}
\end{figure*}

\subsubsection{Temporal Hyperparameter Effects}
\label{sec:temporal_hyperparameters}

Figure~\ref{fig:rastream_results} reports the effect of finite unroll horizon $T$ and streaming stride $s$ on accuracy and smoothness. All configurations use the pretrained Base spatial backbone.
The best configuration, $T=16,s=2$, achieves 72.05 mm MVE on ALL, reducing the RaStream (spatial-only) reference by 12.22 mm (14.5\%). Reductions in MRE and TE show that the causal temporal refiner improves articulated rotation and global localization, not only dense vertex error. We therefore emphasize MVE, MRE, TE, and temporal smoothness in Fig.~\ref{fig:rastream_results}.
Figure~\ref{fig:rastream_results}(b) reports TA and JA with stride-matched RaStream (spatial-only) references for $s=1$ and $s=2$, avoiding cross-stride smoothness comparisons caused by different sampled trajectories.
Among the tested configurations, $T=16,s=2$ obtains the best MVE, MRE, and TE. Increasing $T$ alone does not guarantee the same improvement: the $T=16,s=1$ setting improves rotation error but does not achieve the lowest MVE. The results support selecting $T$ and $s$ jointly.

\begin{figure}[!t]
\centering
\includegraphics[width=\columnwidth]{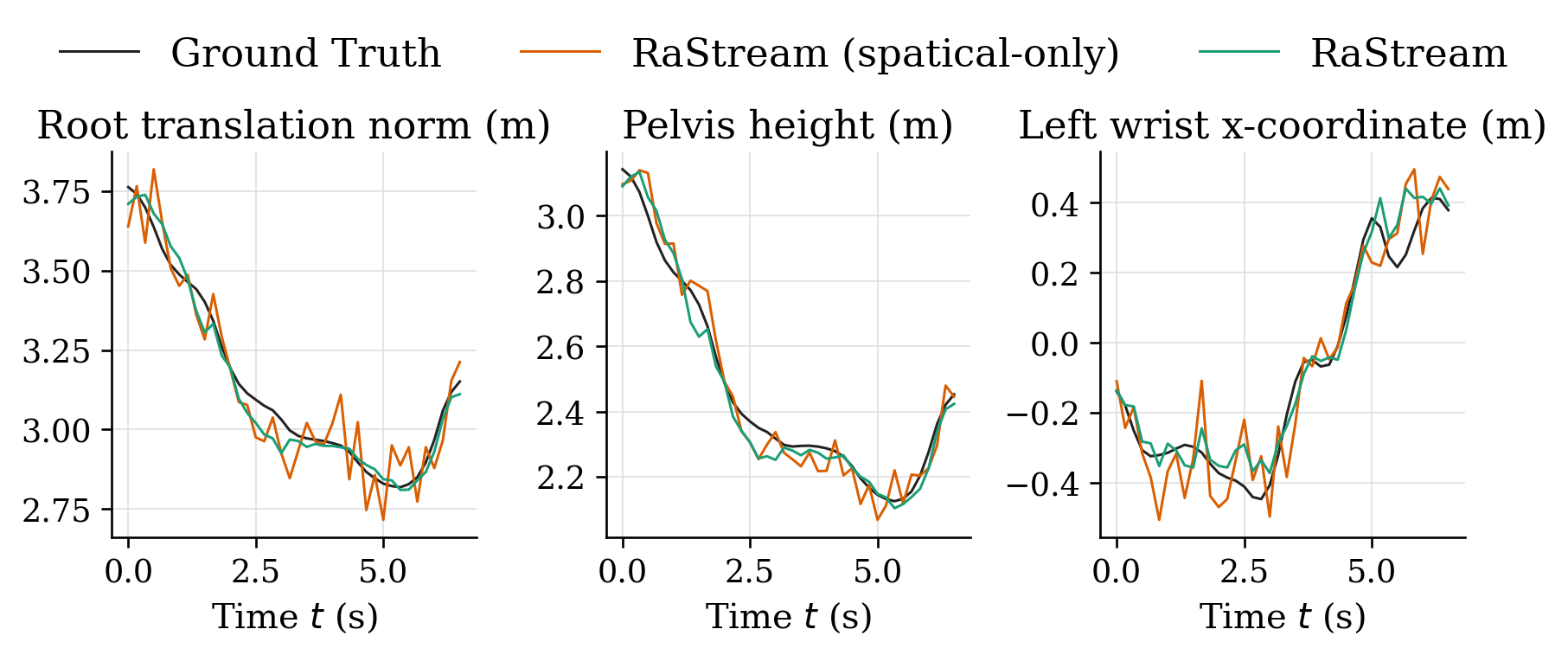}
\caption{Representative temporal trajectory comparison. The horizontal axis denotes time $t$. RaStream produces smoother and more temporally consistent trajectories than the RaStream (spatial-only) variant while preserving the main motion trend.}
\label{fig:temporal_three_panel}
\end{figure}

Figure~\ref{fig:temporal_three_panel} visualizes representative temporal trajectories for root translation, pelvis height, and left-wrist motion. Compared with the RaStream (spatial-only) variant, the complete RaStream framework follows the ground-truth trends with fewer abrupt local fluctuations. The trajectories complement the quantitative TA and JA results.

\subsection{Ablation Studies}
\label{sec:ablations}

\paragraph{Spatial encoder components}
Table~\ref{tab:spatial_component_ablation} uses the single-window protocol. For
``w/o 3D structure,'' the third radar axis is folded into channels and 3D
convolutions are replaced by 2D counterparts. For ``w/o localization and
RoI,'' the detection branch and crop are removed, and the resized full volume
is directly processed by the mesh encoder.
Both variants degrade all metrics, increasing MVE from 84.3 mm to 89.6 mm and
92.1 mm, respectively. This confirms the value of native 3D structure and
localization-conditioned reconstruction.

\begin{table}[t]
\caption{Ablation of the RaStream spatial encoder on the M4Human single-window protocol.}
\label{tab:spatial_component_ablation}
\centering
\footnotesize
\setlength{\tabcolsep}{3.0pt}
\begin{tabular}{lcccc}
\toprule
Variant & MVE$\downarrow$ & MJE$\downarrow$ & MRE$\downarrow$ & TE$\downarrow$ \\
\midrule
Spatial encoder & 84.3 & 71.2 & 9.0 & 45.4 \\
w/o 3D structure & 89.6 & 76.8 & 9.6 & 47.1 \\
w/o localization and RoI & 92.1 & 78.3 & 9.9 & 47.1 \\
\bottomrule
\end{tabular}
\end{table}

\paragraph{Temporal loss components}
Table~\ref{tab:ablation_loss} reports the temporal supervision and temporal architecture ablations. All rows use $T=8,s=1$ and the best validation checkpoint.
Adding all-frame supervision and velocity losses improves the last-frame-only baseline, while translation acceleration gives the lowest MVE and TE in this study.

\begin{table}[t]
\caption{Ablation study on temporal losses and architecture.}
\label{tab:ablation_loss}
\centering
\footnotesize
\setlength{\tabcolsep}{1.2pt}
\begin{tabular*}{\columnwidth}{@{\extracolsep{\fill}}l|ccc|cc@{}}
\toprule
\multirow{2}{*}{Setting} & \multicolumn{3}{c|}{Spatial Accuracy} & \multicolumn{2}{c}{Temporal Smoothness $\downarrow$} \\
\cmidrule(lr){2-4} \cmidrule(lr){5-6}
& MVE & MRE & TE & TA & JA \\
\midrule
\multicolumn{6}{c}{\textit{Temporal loss components}} \\
\midrule
Last frame & 76.30 & 7.68 & 46.44 & 4.64 & 10032.12 \\
All-frame+vel. & 74.59 & \textbf{7.56} & 43.92 & 4.42 & \textbf{9206.04} \\
All-frame+vel.+acc. & \textbf{72.71} & 7.57 & \textbf{42.27} & \textbf{4.39} & 9208.81 \\
\midrule
\multicolumn{6}{c}{\textit{Temporal architecture}} \\
\midrule
Single-state GRU & 74.10 & 7.62 & 43.58 & 4.51 & 9360.42 \\
Dual-state + fixed EMA & 72.89 & 7.57 & 42.44 & \textbf{4.39} & 9235.62 \\
Dual-state + learned EMA & \textbf{72.71} & \textbf{7.57} & \textbf{42.27} & \textbf{4.39} & \textbf{9208.81} \\
\bottomrule
\end{tabular*}
\end{table}

\paragraph{Temporal architecture}
The architecture rows use the same objective and comparable hidden dimensions.
The dual-state model improves over the single-state GRU from 74.10 to 72.71 mm
MVE and from 4.51 to 4.39 TA. Fixed EMA slightly degrades performance,
indicating that state decomposition provides the main gain while adaptive
morphology updating supplies a modest further benefit.

\subsection{Edge Profiling and Pareto Analysis}
\label{sec:edge_pareto}

\subsubsection{Edge Performance Profiling}
\label{sec:edge_profiling}

Table~\ref{tab:deployment} reports FP32 Jetson Orin Nano profiling for representative RaStream configurations using $T=16,s=2$ across the three backbone scales. The MVE column uses the same temporal configuration as the profiled edge setting, while latency, memory, and power characterize the per-invocation streaming path before stride amortization.

\begin{table}[t]
\caption{Measured Jetson Orin Nano FP32 profiling of representative RaStream scale variants with $T=16,s=2$.}
\label{tab:deployment}
\centering
\footnotesize
\setlength{\tabcolsep}{3.0pt}
\begin{tabular}{lccccc}
\toprule
Scale & MVE$\downarrow$ & Latency$\downarrow$ & P95$\downarrow$ & Memory & Power \\
      & (mm) & (ms) & (ms) & (MB) & (idle/active W) \\
\midrule
Tiny  & 77.58 & \textbf{19.69} & \textbf{30.86} & \textbf{201} & 3.9 / 6.8 \\
Small & 75.12 & 23.30 & 33.52 & 240 & 4.3 / 7.2 \\
Base  & \textbf{72.05} & 26.93 & 32.88 & 291 & 4.6 / 7.3 \\
\bottomrule
\end{tabular}
\end{table}

\subsubsection{Pareto Frontier Analysis}
\label{sec:pareto_frontier}

\begin{figure}[t]
\centering
\includegraphics[width=\columnwidth]{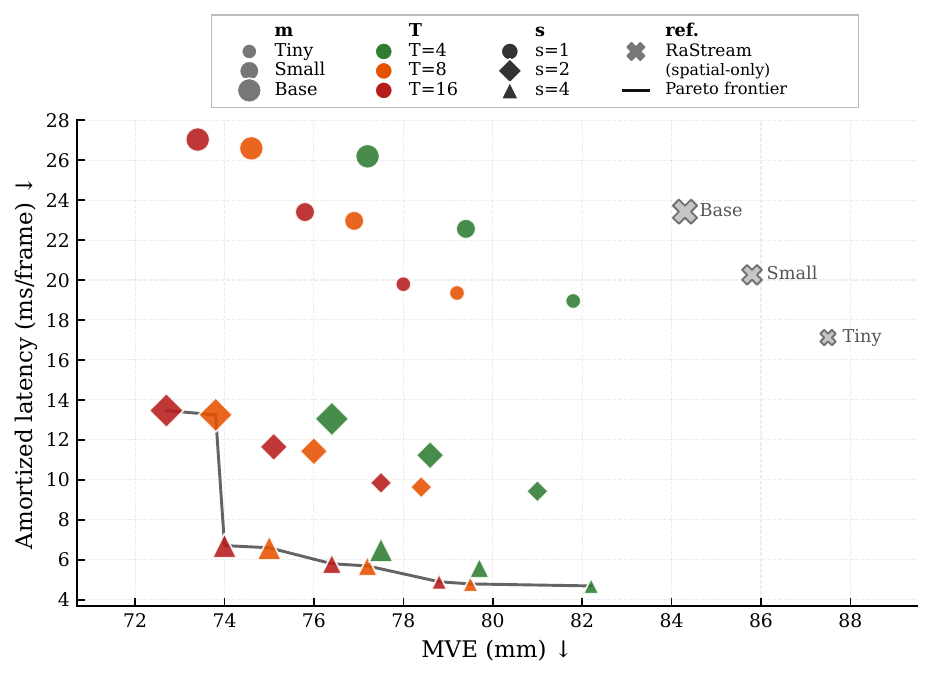}
\caption{Pareto frontier of deployment configurations over reconstruction accuracy and amortized latency. RaStream candidates enumerate the FP32 design space with $T_w=4$, $m\in\{\text{Tiny},\text{Small},\text{Base}\}$, $T\in\{4,8,16\}$, and $s\in\{1,2,4\}$. Marker size denotes model scale, color denotes finite unroll horizon $T$, and marker shape denotes stride $s$. Gray points are RaStream (spatial-only) references.}
\label{fig:scale_latency_tradeoff}
\end{figure}

Figure~\ref{fig:scale_latency_tradeoff} shows the accuracy--latency Pareto
frontier over model scale, unroll horizon, and stride. Larger models favor
reconstruction accuracy, whereas larger strides reduce amortized latency by
lowering spatial-encoder invocation frequency. The Base configuration requires
26.93 ms per invocation, below the 83.3 ms radar-frame interval at 12 Hz.
Thus, RaStream exposes multiple operating points without architectural changes.

\section{Conclusion}
\label{sec:conclusion}

We presented RaStream, a causal streaming framework for mmWave-based SMPL-X
recovery that separates native 3D spatial encoding from compact temporal state
propagation. Its localization-conditioned encoder improves single-window
reconstruction, while causal token-space history aggregation improves accuracy
and temporal consistency. Separating slow morphology and fast motion states
provides an additional gain over single-state recurrence. On M4Human, RaStream
reduces MVE from 90.90 mm for RT-Mesh to 84.27 mm in spatial-only mode and to
72.05 mm with temporal refinement. Jetson Orin Nano profiling demonstrates
practical accuracy--latency trade-offs for edge streaming. Cross-subject and
cross-action generalization remain future work.

\bibliographystyle{IEEEtran}
\bibliography{ref}

\end{document}